\documentclass[english,nofootinbib,aps,prd,twocolumn]{revtex4-1}
\usepackage[T1]{fontenc}
\usepackage[latin9]{inputenc}
\usepackage{color}
\usepackage{amsmath}
\usepackage{amssymb}
\usepackage{graphicx}
\usepackage[percent]{overpic}

\makeatletter

\providecommand{\tabularnewline}{\\}

\@ifundefined{textcolor}{}
{%
 \definecolor{BLACK}{gray}{0}
 \definecolor{WHITE}{gray}{1}
 \definecolor{RED}{rgb}{1,0,0}
 \definecolor{GREEN}{rgb}{0,1,0}
 \definecolor{BLUE}{rgb}{0,0,1}
 \definecolor{CYAN}{cmyk}{1,0,0,0}
 \definecolor{MAGENTA}{cmyk}{0,1,0,0}
 \definecolor{YELLOW}{cmyk}{0,0,1,0}
}

\def\gs{\mathrel{
   \rlap{\raise 0.511ex \hbox{$>$}}{\lower 0.511ex \hbox{$\sim$}}}}
\def\ls{\mathrel{
   \rlap{\raise 0.511ex \hbox{$<$}}{\lower 0.511ex \hbox{$\sim$}}}}

\makeatother

\usepackage{babel}
\begin{document}

\title{ A left-right symmetric flavor symmetry model}

\author{Werner Rodejohann$^a$ and Xun-Jie Xu$^{a,b}$}

\affiliation{$^a$Max-Planck-Institut f\"ur Kernphysik, Postfach 103980, D-69029
Heidelberg, Germany\\
$^b$Institute of Modern Physics and Center for High Energy Physics,
Tsinghua University, Beijing 100084, China}

\date{\today}
\begin{abstract}
\noindent
We discuss flavor symmetries in left-right symmetric theories. 
We show that such frameworks are a different environment 
for flavor symmetry model building compared to the usually considered cases. This 
does not only concern the need to obey the enlarged gauge structure, but also 
more subtle issues with respect to residual symmetries. 
Furthermore, if the discrete left-right symmetry is charge conjugation, potential inconsistencies between the flavor and charge conjugation symmetries should be taken care of. 
In our predictive model based on $A_4$ we analyze the correlations between the 
smallest neutrino mass, the atmospheric mixing angle and the Dirac CP phase, the 
latter prefers to lie around maximal values. There is no lepton flavor violation from the Higgs bi-doublet. 

\end{abstract}
\maketitle
\def\headif{\iftrue}

\headif

\section{Introduction}

Despite the huge and continued success of the Standard Model (SM)
in the last several decades, the flavor structure of the three generations
of fermions in the SM 
leaves a big puzzle that remains to be understood. 
In particular, lepton mixing is so drastically different from quark mixing that 
the field of flavor symmetry model building is among the  busiest 
ones in flavor physics. To avoid Goldstone bosons and to unify 
at least two different generations one typically chooses   
discrete non-Abelian groups as flavor symmetry 
\cite{Altarelli:2010gt,Ishimori:2010au,King:2013eh,Feruglio:2015jfa}. 

Apart from the unusual lepton mixing structure, the second big puzzle introduced by 
neutrino physics is the smallness of neutrino mass. 
An attractive approach is to link this smallness to the parity violation of the SM. 
This is in fact achieved in left-right symmetric models \cite{LR1,LR2,LR3,LR4,Mohapatra:1979ia} 
where the gauge group of the SM is extended to 
$SU(2)_{L}\times SU(2)_{R}\times U(1)_{B-L}$.

Linking the two aspects mentioned so far, we aim in this paper at building a 
flavor symmetry model in a left-right symmetric model (LRSM). 
As Grand Unified Theories (GUTs) based on $SO(10)$ can be 
broken down with an intermediate left-right symmetry to the SM, 
it may be possible to extent such LRSM flavor models 
in a bottom-up strategy to GUT flavor models. 
Our approach could be considered as a first modest step to unify 
particle and chirality species.\\

The constraints that are imposed by left-right symmetry modify some of the well-known 
features of usually considered flavor symmetry models. For instance, 
a typical example \cite{Altarelli:2005yx} based on the most often used flavor 
group $A_4$, assigns the left-handed lepton $SU(2)_L$ doublets as well as 
the right-handed neutrinos to the three-dimensional irreducible representation of $A_4$. 
Right-handed charged fermions instead transform as the three different
one-dimensional representations. This is incompatible with the fact that 
right-handed neutrinos and charged fermions are part of the same gauge doublet. 
In general, models that unify the different particle 
or chirality species are rarely considered and are in general challenging to construct.  

Another issue concerns residual symmetries.  
Usually a discrete flavor symmetry group $G$ is broken to two subgroups $G_{\ell}$ 
and $G_{\nu}$ which constrain the form of the mass matrices $M^{\ell}$
and $M^{\nu}$ for charged leptons and neutrinos respectively. The
mixing matrix is thus essentially determined by the symmetry group. 
The lepton mixing is then independent of the neutrino masses. 
In the minimal left-right symmetric models under study, however, typically 
this direct correlation of subgroups with lepton mixing 
does not exist. The reason is that the neutrino Dirac and the charged lepton mass 
matrices contain in general two contributions as a consequence of the 
Higgs bi-doublet. As a result, even though there are 
in principle conserved subgroups of the flavor group, they do not translate in 
invariance of the mass matrices. 
Therefore lepton mixing will depend on neutrino masses. 
Another issue concerns the discrete left-right symmetry in such models. If it is charge conjugation, one may encounter (depending on the chosen flavor symmetry group) potential inconsistencies between this discrete symmetry and the flavor symmetry. This is then 
similar to the situation when flavor and CP symmetries are combined, see e.g.\ Ref.\ \cite{Holthausen:2012dk}. 

In this paper we will construct a flavor symmetry model based on $A_4$ within a left-right symmetric context. We discuss carefully the general and specific model building aspects of such scenarios and analyze several predictive solutions for the neutrino sector. We show that flavor changing currents in the lepton sector generated by the Higgs bi-doublet are absent. \\

The paper is organized as follow. In Sec.\ \ref{sec:General-flavor-structure} 
we discuss left-right symmetric models and outline aspects of their impact on 
flavor symmetry model building. In Sec.\ \ref{sec:model} we present a 
model based on $A_4$ that is 
compatible with left-right symmetry and analyze it numerically and 
analytically in Sec.\ \ref{sec:result}, in order to demonstrate that it  
is compatible with current data.  
We conclude in Sec.\ \ref{sec:Conclusion}, 
some analytical details are delegated to the Appendix. 


\section{\label{sec:General-flavor-structure}The impact of LRSM on flavor
symmetry }

In this section we first review the aspects of minimal left-right symmetric models (LRSM) that 
we need in this paper and then discuss their impact on building flavor symmetry models.

\subsection{The minimal LRSM}

In the minimal LRSM \cite{Mohapatra:1979ia,LR1,LR2,LR3,LR4} the gauge 
group is $SU(2)_{L}\times SU(2)_{R}\times U(1)_{B-L}$. Right- and
left-handed leptons $\ell_{R}$, $\ell_{L}$ are doublets under $SU(2)_{R}$
and $SU(2)_{L}$ respectively. Three Higgs multiplets 
$\Delta_{L}\sim(3,1,2)$, $\Delta_{R}\sim(1,3,2)$ and $\Phi\sim(2,2,0)$ are introduced
to break $SU(2)_{L}\times SU(2)_{R}\times U(1)_{B-L}$ to $SU(2)_{L}\times U(1)_{Y}$ and 
further to $U(1)_{\rm em}$, respectively. We choose here the left-right parity transformation as
\begin{equation}
\ell_{L}\leftrightarrow\ell_{R}\,,~\Phi\leftrightarrow\Phi^{\dagger}
\,,~\Delta_{L}\leftrightarrow\Delta_{R}\,.\label{eq:0606}
\end{equation}
The Yukawa interactions of the lepton sector are 
\begin{eqnarray}\label{eq:P}
\mathcal{L} & \supset & Y_{ij}\bar{\ell}_{Li}\Phi\ell_{Rj}+\tilde{Y}_{ij}\bar{\ell}_{Li}\tilde{\Phi}\ell_{Rj}\nonumber \\
 & + & (Y_{Lij}\ell_{Li}^{T}\Delta_{L}\ell_{Lj}+Y_{Rij}\ell_{Ri}^{T}\Delta_{R}\ell_{Rj})+h.c.\label{eq:0429}
\end{eqnarray}
The above discrete left-right symmetry leads to $Y = Y^\dagger$, $\tilde Y = \tilde Y^\dagger$ and 
$Y_{L} = Y_{R}$. This can be seen in particular by comparing the term 
$Y_{ij}\bar{\ell}_{Li}\Phi\ell_{Rj}$ and its hermitian conjugate 
$(Y^\dagger)_{ij}\bar{\ell}_{Ri}\Phi^\dagger \ell_{Lj}$ with the 
parity-transformed terms $Y_{ij}\bar{\ell}_{Ri}\Phi^\dagger\ell_{Lj}$ 
and $(Y^\dagger)_{ij}\bar{\ell}_{Li}\Phi\ell_{Rj}$. 

The scalar fields acquire the following vacuum expectation values 
\begin{equation}\nonumber
\langle\Phi\rangle=\left(\begin{array}{cc}
\kappa & 0\\
0 & \kappa'
\end{array}\right),~\langle\Delta_{L}\rangle=(0,0,v_{L})\,,~\langle\Delta_{R}\rangle=(0,0,v_{R})\,.\label{eq:0429-1}
\end{equation}
From now on we will assume that $v_L$ is sufficiently small to be neglected. 
The neutrino Dirac mass matrix $m_D$ and the charged lepton mass matrix 
$M^{\ell}$ are given as 
\begin{equation}
m_{D}  =\kappa Y+\kappa'\tilde{Y} \,,~M^{\ell}  =\kappa'Y+\kappa\tilde{Y}\,.
\label{eq:0602}
\end{equation}
which implies that for given $m_{D}$ and $M^{\ell}$ one can always 
find the associated $Y$ and $\tilde{Y}$ as long as $\kappa^{2}\neq\left(\kappa'\right)^{2}$. 
The relative contribution to the mass matrices is determined 
by the ratio
\begin{equation}\label{eq:tanb}
\tan\beta\equiv\kappa/\kappa'\,.
\end{equation}
The right-handed neutrinos have a Majorana mass matrix
\begin{equation}
M_{R}=v_{R}Y_{R}\label{eq:0429-4}\,,
\end{equation}
which generates the light neutrino masses via the type I seesaw 
\begin{equation}
M^{\nu}=-m_{D}M_{R}^{-1}m_{D}^{T}\,.\label{eq:0512-14}
\end{equation}
With the simple and straightforward assumption of $m_D$ lying around the weak scale, 
$M_R$ lies around $10^{15}$ GeV, which implies that the scale of parity restoration and thus also the right-handed gauge boson masses lie around that scale. 

\subsection{Left-right symmetry and flavor symmetries\label{sec:confusing}}

We mention here some aspects that are connected to left-right symmetry and flavor symmetry model building. We focus on $A_4$ here, but our statements will 
hold for many other groups as well. 

Note first that the left- and right-handed 
lepton doublets, as well as the left- and right-handed Higgs triplets have to 
transform in the same representation of the flavor symmetry group. As 
right-handed fermions live in a gauge group doublet now, the right-handed neutrinos and 
the charged fermions of a given generation transform together. 
This means that popular $A_4$ models with the left-handed doublets as triplet and 
the right-handed charged fermions as singlets are not possible. Also models 
in which the right-handed neutrinos transform as triplet and the 
right-handed charged fermions as singlets are forbidden. \\

In typical flavor symmetry models, the Yukawa terms are effective in the sense that apart from Higgs, left- and right-handed fermion fields in addition 
scalar flavon fields are present. The full Yukawa term (keeping the bi-doublet $\Phi$ as trivial singlet of the flavor group) can be written in the usual compact form as 
\begin{equation}\label{eq:yukLRfla}
Y_{ij}\bar{\ell}_{Li}\Phi\ell_{Rj}\phi\,,
\end{equation}
where $\phi$ is the flavon field. If $\ell_{L,R}$ are multiplets and $\phi$ 
is a trivial singlet of the flavor group, then as usual $Y = Y^\dagger$.  
Consider now the case when $\ell_{L,R}$ and 
$\phi$ are non-trivial multiplets of the flavor group. In this case 
$Y_{ij}\bar{\ell}_{Li}\Phi\ell_{Rj}\phi$ should be written as 
$\sum_k Y_{ij}^k\bar{\ell}_{Li}\Phi\ell_{Rj}\phi^k$, which means that there will be several 
Yukawa coupling matrices. For instance, in $A_4$ the full Yukawa term could be a triple-triplet term, i.e.\ $\ell_{L,R}$ and $\phi$ are all triplets. Then, because the product of two triplets 
contains two triplets according to $3\times3=3+3+1+1+1$, we have two different Yukawa matrices $Y_1$ and $Y_2$. Following the steps as given after Eq.\ (\ref{eq:0429}), one 
finds that 
\begin{equation}\label{eq:hermLR}
\sum_k Y_k = \left(\sum _kY_k \right)^\dagger\,.
\end{equation}
Actually we have here assumed real flavon fields, but the same results applies for 
complex fields. 
As a physical result of Eq.\ (\ref{eq:hermLR}), the PMNS matrix of the left-handed 
leptons will be equal to its right-handed analog.

We also note that the definition of the discrete left-right symmetry is not unique 
in LR symmetric models. One could also choose charge conjugation, which would replace 
in Eq.\ (\ref{eq:hermLR}) the $^\dagger$ with $^T$. However, this choice of 
discrete left-right symmetry would bring along the complications that the 
flavor symmetry group transformations are potentially incompatible with 
the charge conjugation, similar to the situation of combining flavor symmetry with 
CP symmetry, see e.g.\ Ref.\ \cite{Holthausen:2012dk}. In particular, 
for different flavor groups one would need to introduce different non-trivial 
charge conjugations in the LRSM. In this paper we only focus on 
parity as discrete left-right symmetry, leading to Eq.\ (\ref{eq:hermLR}). 
In more general models with different definitions of the 
discrete left-right symmetry a careful check of the consistency would need to be performed.

Another point we wish to make concerns residual symmetries. 
Typical models break $A_4$ in such a way that in the neutrino and 
charged lepton sector subgroups of $A_4$ remain intact\footnote{Sometimes 
those residual symmetries are also accidental.}. 
In general, a flavor group $G$ breaks to different subgroups 
$G_\ell$ and $G_\nu$ in the charged lepton and neutrino sector, respectively: 
\begin{equation}
G\rightarrow\begin{cases}
G_{\ell}: & \{T|\thinspace T^{\dagger}M^{\ell}M^{\ell\dagger}T=M^{\ell}M^{\ell\dagger}\}\\
G_{\nu}: & \{S|\thinspace S^{T}M^{\nu}S=M^{\nu}\}
\end{cases}.\label{eq:1210-5}
\end{equation}
The eigenvectors of $T$ are just the columns of 
the mixing matrix $U_{\ell}$ which diagonalizes the charged lepton sector,
and likewise in the neutrino sector $S$ determines $U_{\nu}$. Thus,  
the PMNS matrix given by $U_{\ell}^{\dagger}U_{\nu}$ is essentially
determined by $G_{\nu}$, $G_{\ell}$, 
irrespective of the dynamical realization within a 
model \cite{Lam:2008rs,Lam:2008sh,Lam:2011ag}. This implies in particular 
that mixing is independent of masses. 
It is thus possible to reconstruct the flavor group $G$ from the 
mixing matrix $U$, or vice versa to break $G$ into proper subgroups 
to obtain $U$. Both the $U\Rightarrow G$ 
and $G\Rightarrow U$ procedures have been well understood and there 
are many studies on this subject 
\cite{Holthausen:2013vba,Araki:2013rkf,King:2013vna,Hernandez:2012sk,He:2012yt,He:2015xha,Ge:2011qn,Rodejohann:2015pea,Toorop:2011re,Toorop:2011jn,Fonseca:2014koa,Grimus:2011fk,Toorop:2011re,Toorop:2011jn}. 
If in a given model with a seesaw mechanism the right-handed 
Majorana mass matrix is assumed to be proportional to 
the unit matrix, or if 
$m_{D}$ and $M_{R}$ share the same residual symmetry $G_{\nu}$ 
(hence can be diagonalized simultaneously), 
the above game can again be played and with identifying 
the residual symmetries of $M^\nu$ and 
$M_\ell$, information on the original flavor symmetry group could be obtained.

What concerns left-right symmetric models is that the 
Dirac and charged lepton mass matrices are given by contributions of two 
fundamental terms, $Y$ and $\tilde Y$, see Eq.\ (\ref{eq:0602}). Their relative 
contribution is governed by $\tan \beta$ in Eq.\ (\ref{eq:tanb}). 
Only in the limit $\tan\beta\rightarrow\infty$ the minimal LR model 
is similar to the SM, as in this case only $Y$ contributes to Dirac 
neutrino masses 
and $\tilde{Y}$ to charged lepton masses. 
In this limit of $\kappa'\ll\kappa$ the symmetry of $m_{D}$ is the one of 
$Y$. Once $\kappa'/\kappa$ is non-zero $m_{D}$ has neither the symmetry of 
$Y$ nor of $\tilde{Y}$. Similar statements hold for 
$\tan\beta\rightarrow 0$. 

In left-right symmetric models 
$m_{D}$ and $M_{R}$ cannot share the same residual symmetry $G_{\nu}$ 
and hence cannot be diagonalized simultaneously:  
the fact that in Eq.\ (\ref{eq:0602}) two contributions to 
$M_\ell$ and $m_D$ are present, means  
that there is no non-trivial symmetry basis in which this can happen, 
unless $\tan\beta\rightarrow\infty$ or $\tan\beta\rightarrow 0$. 

If neutrino mass would be given by a dominating type II 
seesaw term, i.e.\ the contribution of type I seesaw which involves $m_D$ is suppressed, 
then in principle the residual symmetries can be well separated. 


One can therefore conclude: if we introduce a flavor group and 
intend to break it into two parts for neutrinos and charged leptons respectively, 
then within left-right symmetric models this is impossible 
unless $\tan \beta$ takes on extreme values or the contribution 
of type I seesaw to neutrino masses is absent. If this is not the case, 
the simple connection between the flavor symmetry subgroups and $U$ no 
longer applies. To put it in another way, if some VEV alignment would lead 
to simple residual symmetries and a 
simple mixing structure in a model without left-right symmetry, 
the presence of a left-right symmetry leads to deviations. 

As is well known, the presence of the Higgs bi-doublet and thus two Dirac Yukawa 
contributions in Eq.\ (\ref{eq:0602}) implies potentially dangerously rates for 
lepton flavor violation (LFV), see \cite{Barry:2013xxa} for a compliation. 
While the Higgs triplets and processes involving the right-handed 
gauge bosons and neutrinos also lead to LFV, their contributions are naturally suppressed if the scale of parity restoration lies above, say, 10 TeV. This is in fact expected from simple neutrino mass constraints, where the mass scale of the right-handed neutrinos is almost GUT scale, see 
Eq.\ (\ref{eq:0512-14}).  
Already in the very early Ref.\ 
\cite{Mohapatra:1979ia} the dangerous LFV generated by the bi-doublet was noted and taken care of by imposing a simple $Z_2$  symmetry to suppress 
$\mu\rightarrow e \gamma$ and $\mu\rightarrow 3e$. Hence, a 
flavor symmetry can be very useful and important in order to avoid LFV. 
Generally speaking, if  $Y$ and $\tilde{Y}$ in Eqs.\ (\ref{eq:0429},\,\ref{eq:0602}) cannot be 
simultaneously diagonalized, LFV processes generated by the bi-doublet Dirac Yukawas are not suppressed. 
If $Y$ and $\tilde{Y}$ can be made simultaneously diagonal, such processes are absent. 
As we will see in the next Section, our model has this feature. 


\fi

\headif

\section{\label{sec:model}$A_{4}$-LRSM model}

\begin{table}
\centering

\begin{tabular}{|c|c|c|c|c|c|}
\hline 
 & $A_{4}$ & $SU(2)_{L}$ & $SU(2)_{R}$ & $U(1)_{B-L}$ & $Z_{2}$\tabularnewline
\hline 
\hline 
$\ell_{L}$ & $3$ & $2$ & $1$ & $-1$ & $0$\tabularnewline
\hline 
$\ell_{R}$ & $3$ & $1$ & $2$ & $-1$ & $0$\tabularnewline
\hline 
$\Phi$ & $1$ & $2$ & $2$ & $0$ & $1$\tabularnewline
\hline 
$\phi^{\ell}$ & $3$ & $1$ & $1$ & $0$ & $1$\tabularnewline
\hline 
$\phi^{\nu}$ & $3$ & $1$ & $1$ & $0$ & $0$\tabularnewline
\hline 
$\xi$ & $1$ & $1$ & $1$ & $0$ & $1$\tabularnewline
\hline 
$\Delta_{L}$ & $1$ & $3$ & $1$ & $2$ & $0$\tabularnewline
\hline 
$\Delta_{R}$ & $1$ & $1$ & $3$ & $2$ & $0$\tabularnewline
\hline 
\end{tabular}

\protect\caption{\label{tab:Particles}Particle content of the model.}
\end{table}

The flavor symmetry in this model is $A_{4}\times Z_{2}$ and the 
particle content with its transformation properties 
is given in Tab.\ \ref{tab:Particles}. Note that the left- and right-handed 
lepton doublets, as well as the left- and right-handed Higgs triplets 
transform in identical representation of the flavor symmetry group. 
In addition to the standard LRSM particles we only introduce two $A_{4}$ triplets ($\phi^{\ell}$, 
$\phi^{\nu}$) and one $A_{4}$ singlet $\xi$. 
The Lagrangian of all Yukawa interactions can be written as 
\begin{eqnarray}\nonumber 
\mathcal{L} & \supset & \bar{\ell}_{L}(Y_{\xi}\xi+Y_{\ell1}\phi^{\ell}+Y_{\ell2}\phi^{\ell})\Phi\ell_{R}\label{eq:0512-1}\\ \nonumber 
 & + & \bar{\ell}_{L}(\tilde{Y}_{\xi}\xi+\tilde{Y}_{\ell1}\phi^{\ell}+\tilde{Y}_{\ell2}\phi^{\ell})\tilde{\Phi}\ell_{R}\\\nonumber 
 & + & \ell_{R}^{T}(Y_{R}^{0}+Y_{R}^{\nu}\phi^{\nu})\Delta_{R}\ell_{R}\\
 & + & \ell_{L}^{T}(Y_{R}^{0}+Y_{R}^{\nu}\phi^{\nu})\Delta_{L}\ell_{L}\,.
\end{eqnarray}
Note the presence of two terms with 
$\bar \ell_L \phi^\ell \ell_R$, as it is a triple-triplet product, see the discussion around 
Eq.\ (\ref{eq:yukLRfla}). 
For simplicity, we suppress all flavor indices in the Lagrangian. 
Choosing for convenience the real 3-dimensional representation of $A_{4}$, 
it follows that $Y_{\xi}$, $\tilde{Y}_{\xi}$ and $Y_{R}^{0}$ are proportional 
to the unit matrix. The terms involving $Y_{\ell1}$ are 
governed by 
\begin{equation}
y \left(\begin{array}{ccc}
0 & \phi_{3}^{\ell} & 0\\
0 & 0 & \phi_{1}^{\ell}\\
\phi_{2}^{\ell} & 0 & 0
\end{array}\right). \label{eq:0512-3}
\end{equation}
Identical flavor structure holds for $\tilde{Y}_{\ell1}$. 
The terms involving $Y_{\ell 2}$ are, obeying the consistency relation from 
Eq.\ (\ref{eq:hermLR}) proportional to 
\begin{equation}
y^\ast \left(\begin{array}{ccc}
0 & 0 & \phi_{2}^{\ell}\\
\phi_{3}^{\ell} & 0 & 0\\
0 & \phi_{1}^{\ell} & 0
\end{array}\right)\label{eq:0512-5}
\end{equation}
with again identical flavor structure of $\tilde{Y}_{\ell2}$. 
We assume here symmetry breaking of the flavor symmetry according to
the usual vacuum expectation value alignment  
\begin{equation}
\langle\phi^{\ell}\rangle\propto(1,1,1)\,,\thinspace\langle\phi^{\nu}\rangle\propto(0,1,0)\,.
\label{eq:0512-6}
\end{equation}
Combining $Y_\xi$ with the structure of  ${Y}_{\ell1}$ and ${Y}_{\ell2}$ gives 
\begin{equation}
Y=\left(\begin{array}{ccc}
\alpha & \beta & \gamma\\
\gamma & \alpha & \beta\\
\beta & \gamma & \alpha
\end{array}\right)\label{eq:0618-3}
\end{equation}
with the constraint $\alpha = \alpha^\ast$ and $\beta = \gamma^\ast$. Also 
$\tilde Y$ has this structure. Therefore, $Y$ and $\tilde{Y}$ can be simultaneously
diagonalized which implies that the Dirac mass matrices of charged
leptons and neutrinos can be simultaneously diagonalized. Note that this feature implies the absence of potentially dangerous LFV processes generated by the Higgs bi-doublet, 
as discussed at the end of Sec.\  \ref{sec:General-flavor-structure}. 

The remaining symmetric Yukawa matrix resulting from $Y_{R}^{\nu}$ is proportional to 
\begin{equation}
\left(\begin{array}{ccc}
0 & \phi_{3}^{\nu} & \phi_{2}^{\nu}\\
\phi_{3}^{\nu} & 0 & \phi_{1}^{\nu}\\
\phi_{2}^{\nu} & \phi_{1}^{\nu} & 0
\end{array}\right),\label{eq:0512-5}
\end{equation}
leading to 
\begin{equation}
Y_{R}=\left(\begin{array}{ccc}
a & 0 & b\\
0 & a & 0\\
b & 0 & a
\end{array}\right).\label{eq:0512-8}
\end{equation}
Towards an explicit form of the light neutrino mass matrix we first perform the transformation
\begin{equation}
\ell_{L}\rightarrow\ell'_{L}\equiv U_{\rm W}^{\dagger}\ell_{L}\,, 
\thinspace\ell_{R}\rightarrow\ell'_{R}\equiv U_{\rm W}^{\dagger}\ell_{R} \label{eq:0512-9}
\end{equation}
with the Wolfenstein matrix $U_{\rm W}$ (here $\omega = e^{2\pi i/3}$) 
\begin{equation}
U_{\rm W}=\frac{1}{\sqrt{3}}\left(\begin{array}{ccc}
1 & 1 & 1\\
1 & \omega^{2} & \omega\\
1 & \omega & \omega^{2}
\end{array}\right).\label{eq:0409-3-1}
\end{equation}
As a result of this transformation, $Y$, $\tilde{Y}$ and $Y_{R}$ are transformed to 
$Y'$, $\tilde{Y}'$ and $Y'_{R}$ where $Y'$, $\tilde{Y}'$ are diagonal matrices and
\begin{equation}
Y'_{R}=U_{\rm W}^{T}U_{13}\textnormal{diag}(a+b,a,a-b)U_{13}^{T}U_{\rm W}\,.\label{eq:0512-10}
\end{equation}
Inverting this expression,   
\begin{equation}
(Y'_{R})^{-1}=U_{\rm W}^{\dagger}U_{13}\textnormal{diag}(\frac{1}{a+b},\frac{1}{a},\frac{1}{a-b})U_{13}^{T}U_{\rm W}^{*}\label{eq:0512-11}\,.
\end{equation}
Here we have defined the matrix  
\begin{equation}
U_{13}=\left(\begin{array}{ccc}
\frac{1}{\sqrt{2}} & 0 & \frac{-1}{\sqrt{2}}\\
0 & 1 & 0\\
\frac{1}{\sqrt{2}} & 0 & \frac{1}{\sqrt{2}}
\end{array}\right).\label{eq:0409-6-1-1}
\end{equation}
As common in many $A_{4}$ models, $U_{\rm W}^{\dagger}U_{13}$ gives tri-bimaximal 
mixing, to be more specific: 
\begin{equation}
U_{\rm W}^{\dagger}U_{13}=U'U_{\rm TBM}U'' \,,\label{eq:0512-13}
\end{equation}
where $U'=\textnormal{diag}(1,\omega,-\omega^{2})$, $U''=\textnormal{diag}(1,1,i)$
and 
\begin{equation}
U_{\rm TBM}=\left(\begin{array}{ccc}
\sqrt{\frac{2}{3}} & \frac{1}{\sqrt{3}} & 0\\
\frac{-1}{\sqrt{6}} & \frac{1}{\sqrt{3}} & \frac{1}{\sqrt{2}}\\
\frac{1}{\sqrt{6}} & \frac{-1}{\sqrt{3}} & \frac{1}{\sqrt{2}}
\end{array}\right).\label{eq:0512-12}
\end{equation}
Therefore Eq.\ (\ref{eq:0512-11}) can also 
be written as $(Y'_{R})^{-1}\propto U' X_{\rm TBM}U'$,  
where we have defined  
\begin{equation}
X_{\rm TBM}\equiv U_{\rm TBM}\left(\begin{array}{ccc}
\frac{1}{1+z} & 0 & 0\\
0 & 1 & 0\\
0 & 0 & \frac{-1}{1-z}
\end{array}\right)U_{\rm TBM}^{T}\,.\label{eq:0512-15}
\end{equation}
Here $z\equiv b/a$ is in general a complex number. Since in the type I 
seesaw the light neutrino mass matrix is $M^{\nu}=-m_{D} M_{R}^{-1} m_{D}^{T}$, 
where $m_{D}$ is diagonalized with the transformation (\ref{eq:0512-9}), 
we can write $M^{\nu}$ as 
\begin{equation}
M^{\nu}=m\left(\begin{array}{ccc}
1 & 0 & 0\\
0 & r_{2} & 0\\
0 & 0 & r_{3}
\end{array}\right) X_{\rm TBM}\left(\begin{array}{ccc}
1 & 0 & 0\\
0 & r_{2} & 0\\
0 & 0 & r_{3}
\end{array}\right).\label{eq:0512-16}
\end{equation}
Only $m$ 
has the dimension of mass while the other quantities are all dimensionless. 
Note that the re-phasing  $M^{\nu}\rightarrow P M^{\nu} P^{\dagger}$ 
with $P=\textnormal{diag}(e^{i\theta_{1}},e^{i\theta_{2}},e^{i\theta_{3}})$ 
does not have physical meaning so we can always assume $m$ and $r_{2}$, 
$r_{3}$ in Eq.\ (\ref{eq:0512-16}) to be real numbers. 

Finally, we can give the final form of the light neutrino mass matrix in the charged lepton basis:  
\begin{equation}
M^{\nu}=\frac{m}{3(1+z)}\left(\begin{array}{ccc}
3+z & zr_{2} & -zr_{3}\\
\cdot & \frac{z(2+z)r_{2}^{2}}{z-1} & \frac{\left(3+z-z^{2}\right)r_{2}r_{3}}{z-1}\\
\cdot & \cdot & \frac{z(2+z)r_{3}^{2}}{z-1}
\end{array}\right).\label{eq:0512}
\end{equation}
Note that in the limit $r_2 = r_3 = 1$, $M^{\nu}=m X_{\rm TBM}$ leads to 
TBM and the neutrino mass sum-rule $1/\tilde m_1 - 1/\tilde m_3 = 2/\tilde m_2$ 
(here the masses are understood to be complex, see e.g.\ \cite{Barry:2010yk}) since the 
three neutrino masses are proportional to $1/(1+z),\,1,\,-1/(1-z)$ respectively. 

%

\section{\label{sec:result}Numerical and analytical results}

In our left-right symmetric $A_{4}$ model the light neutrino mass
matrix is given by Eq.\ (\ref{eq:0512}) while the charged leptons are 
diagonal with enough parameters to fully fit their masses. 
First we will numerically diagonalize $M^\nu$ in order to find all 
possible parameter values. Analytical diagonalization of the 
general mass matrix turns out to be rather complicated, so we will 
only give one example. Note that, in the spirit of the discussion in 
Sec.\ \ref{sec:confusing},  the VEV alignment in Eq.\ (\ref{eq:0512-6}) 
breaks $A_4$ to subgroups, but they do not end up in the mass matrices. Hence, the 
mixing will depend on the values of the masses.

\subsection{Numerical solutions}

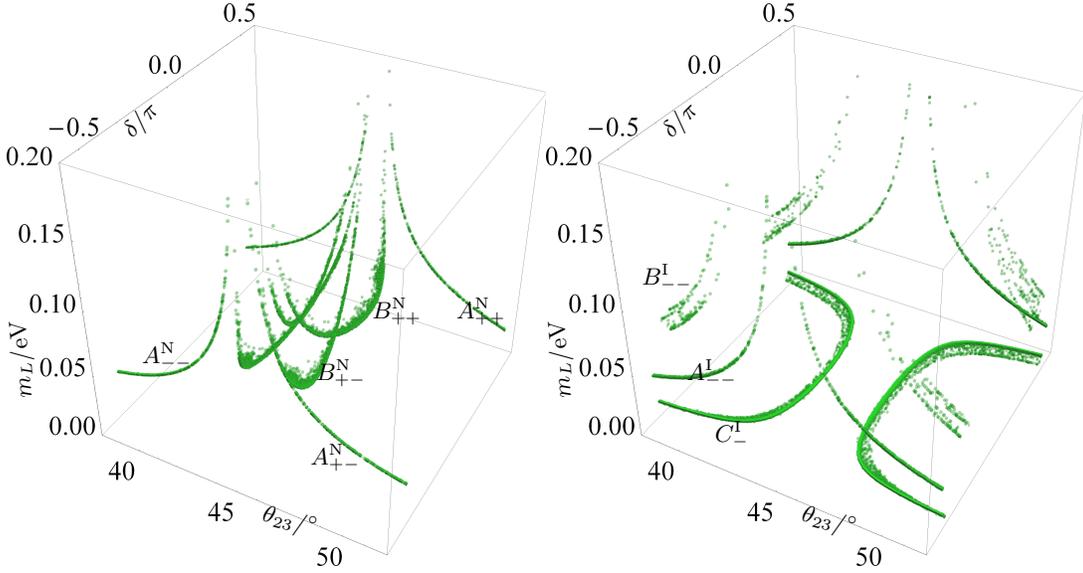
\begin{figure*}
\centering
\begin{overpic}[width=0.4\textwidth
]{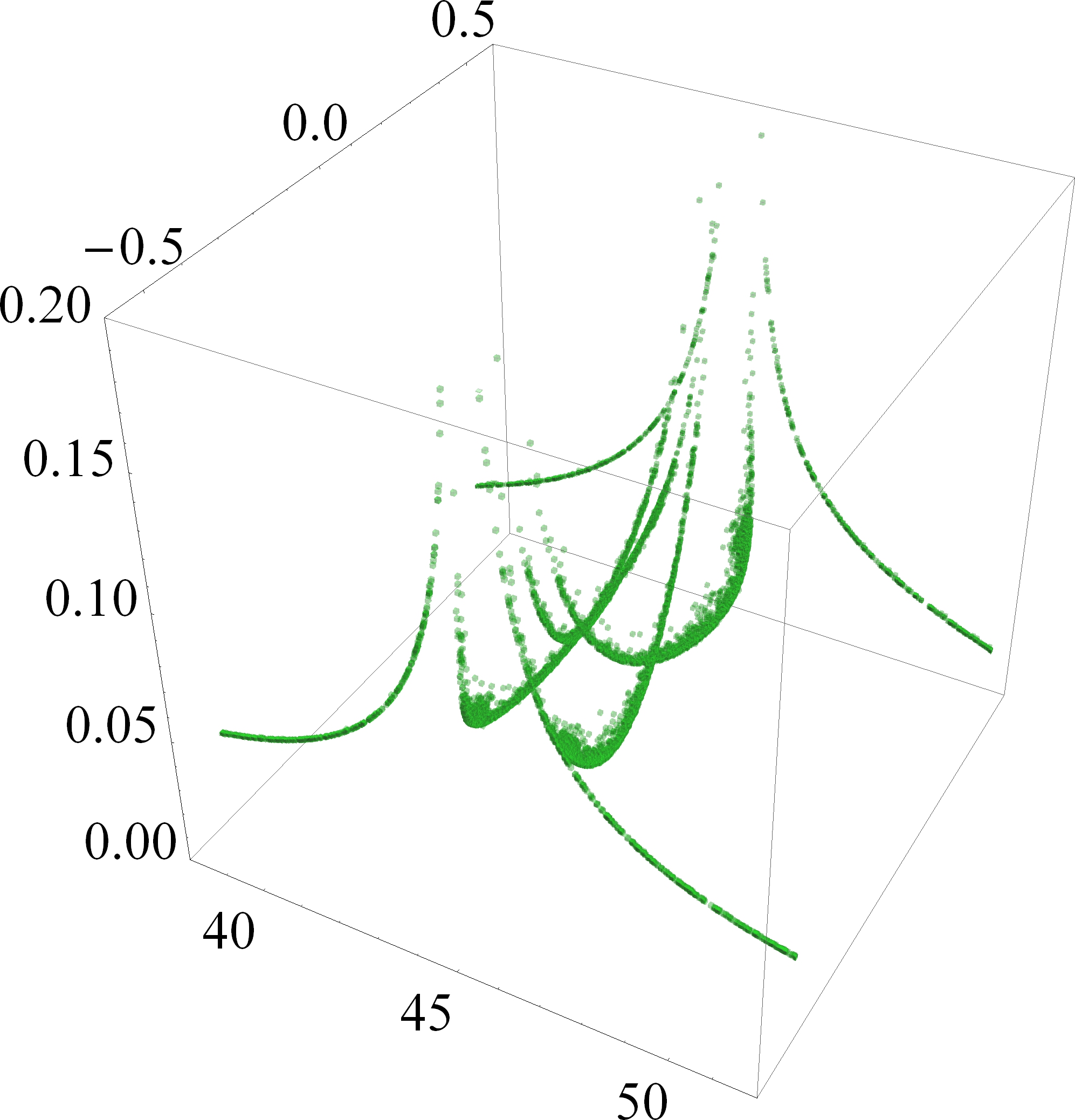}
\put (24,36) {\small$A_{--}^{\rm N}$}
\put (54,18) {\small$A_{+-}^{\rm N}$}
\put (80,44) {\small$A_{++}^{\rm N}$}
\put (65,44) {\small$B_{++}^{\rm N}$}
\put (55,33) {\small$B_{+-}^{\rm N}$}

\put (0,30) {\rotatebox{95}{\small$m_L/\rm{eV}$}}
\put (45,8) {\rotatebox{-15}{\small$\theta_{23}/^{\circ}$}}
\put (20,76) {\rotatebox{45}{\small$\delta/\pi$}}
\end{overpic}\begin{overpic}[width=0.4\textwidth
]{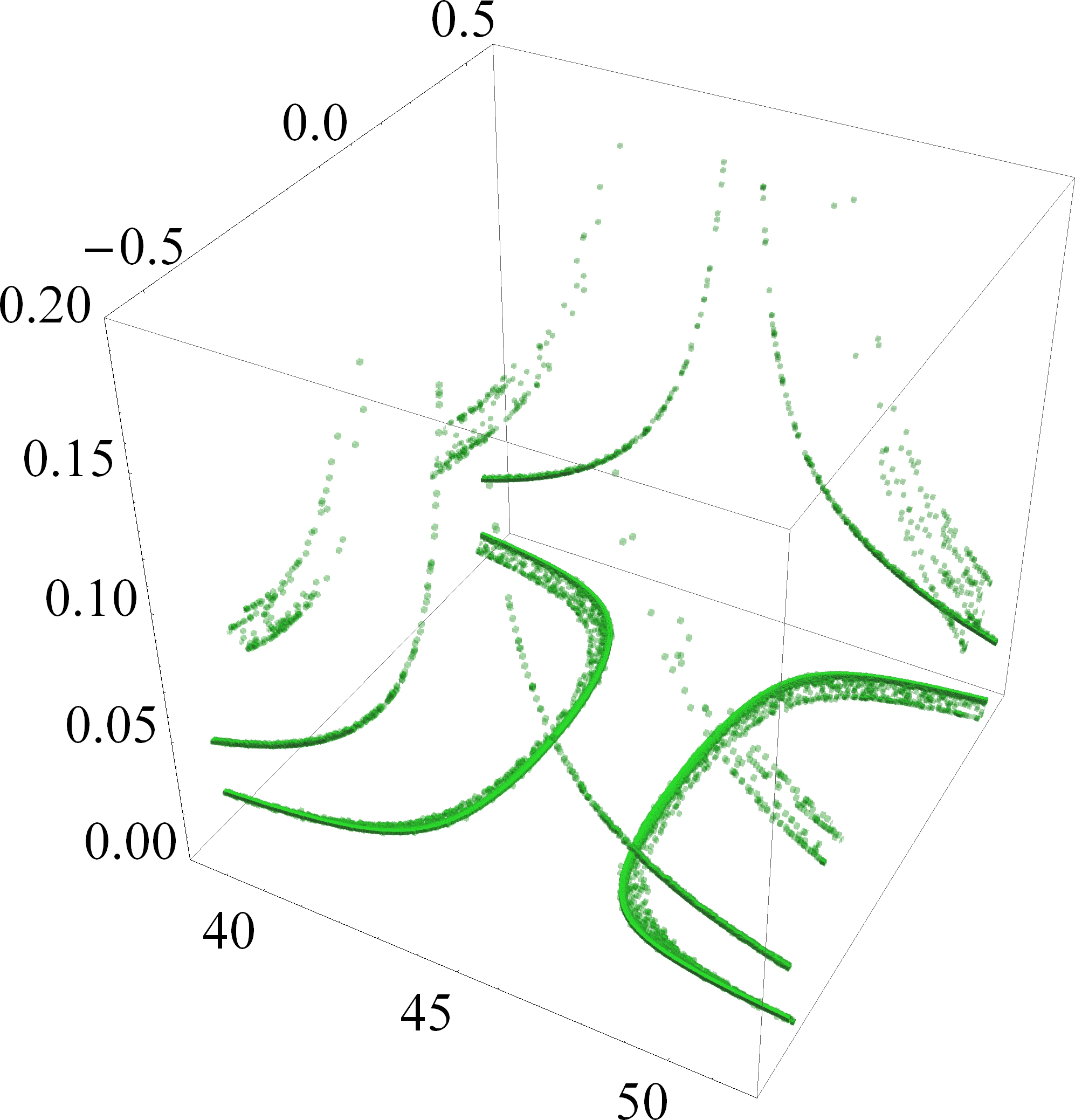}
\put (25,33) {\small$A_{--}^{\rm I}$}
\put (17,50) {\small$B_{--}^{\rm I}$}
\put (30,22) {\small$C_{-}^{\rm I}$}
\put (0,30) {\rotatebox{95}{\small$m_L/\rm{eV}$}}
\put (45,8) {\rotatebox{-15}{\small$\theta_{23}/^{\circ}$}}
\put (20,76) {\rotatebox{45}{\small$\delta/\pi$}}
\end{overpic}
\protect\caption{\label{fig:sol}The $8$ solutions for normal (top) 
and 10 for inverted ordering (bottom) in $\theta_{23}-\delta-m_{L}$
space. 
}
\end{figure*}

Varying all 5 free parameters  $(r_{2},r_{3},m \mbox{ and complex }z)$ 
in Eq.\ (\ref{eq:0512}) and comparing the 
mixing angles and masses with the $3\sigma$ 
global fit results from Ref.\ \cite{NeuFitLisi} reveals that there are 
several disconnected ranges of parameters. The 
eight different cases for the normal ordering and the ten cases 
for the inverted ordering can be seen in  
Fig.\ \ref{fig:sol}, where we plot them in the parameter space 
of $\theta_{23}$, $\delta$ and the smallest mass $m_L$.  
Note that some solutions overlap, but this happens only 
because of the three-dimensional plot. The space of solutions 
is actually five-dimensional 
and the areas in that parameter space do not overlap.

\begin{table*}
\begin{tabular}{|c|c|c|c|c|c|c|}
\hline 
 & \multicolumn{1}{c}{} & \multicolumn{1}{c}{input parameters} &  & \multicolumn{1}{c}{} & \multicolumn{1}{c}{output parameters} & \tabularnewline
\cline{2-7} 
type & $r_{2},\thinspace r_{3}$ & $z$ & $m/\textrm{eV}$ & $(\theta_{23},\theta_{13},\theta_{12},\delta)/^{\circ}$ & $(10^{5}\delta m^{2},10^{3}\Delta m^{2})/\textrm{eV}^{2}$ & $m_{L}/\textrm{eV}$\tabularnewline
\hline 
$A^{\rm N}$ & $-0.439428,\thinspace2.58285$ & $-0.0409371+0.0171862 i$ & $0.0725532$ & $42.5,\thinspace9.2,\thinspace33.9,\thinspace-88.2$ & $7.46,\thinspace2.42$ & $0.074$\tabularnewline
\hline 
$B^{\rm N}$ & $1.10094,1.16147$ & $-0.152655+0.422545i$ & $0.0625143$ & $43.9,\thinspace9.0,\thinspace33.9,\thinspace-60.7$ & $7.33,\thinspace2.42$ & $0.065$\tabularnewline
\hline 
$A^{\rm I}$ & $-2.27434,\thinspace0.399492$ & $0.0325321-0.0126102i$ & $0.10147$ & $43.1,\thinspace9.2,\thinspace34.7,\thinspace-89.6$ & $7.75,\thinspace-2.42$ & $0.087$\tabularnewline
\hline 
$B^{\rm I}$ & $1.04956,\thinspace-0.947632$ & $-0.00592698-0.144104i$ & $0.223795$ & $42.5,\thinspace9.5,\thinspace33.5,\space80.1$ & $7.33,\thinspace-2.42$ & $0.22$\tabularnewline
\hline 
$C^{\rm I}$ & $0.400877,\thinspace0.369437$ & $0.9466+0.208355i$ & $0.0723235$ & $42.7,\thinspace9.0,\thinspace34.2,\thinspace-57.7$ & $7.31,\thinspace-2.42$ & $0.0079$\tabularnewline
\hline 
\end{tabular}\protect\caption{\label{tab:Examples-of-numerical}Examples of 
numerical solutions. The names of the five types of solutions 
are introduced in the text. 
}
\end{table*}

For the normal ordering there are four curves in the shape 
of a ``J'' and another four in the shape of an ``U''. All require a smallest 
neutrino mass above zero, the ones in U-shape have a larger minimal value than 
the ones of J-shape. We name the solutions $A^{\rm N}_{\pm\pm}$ and 
$B^{\rm N}_{\pm\pm}$. The subscript $\pm\pm$ denotes the signs of 
$\theta_{23} - \pi/4$ and $\delta$ (lying in our convention between $-\pi$ and 
$\pi$). 
Interestingly, solutions of type $A$ have values of the CP phase 
very close to $\pm \pi/2$, where $-\pi/2$ seems to be preferred by 
current data \cite{Elevant:2015ska}. 
The type $A$ solutions always keep the signs of 
$\theta_{23} - \pi/4$ and $\delta$, those of type $B$ 
only for most of the parameter space. 
While the lower limit on the smallest 
mass is 0.034 eV for type $A$, it is 0.046 eV for type $B$. 

There are similar types of solutions for the inverted mass ordering, 
denoted $A^{\rm I}_{\pm\pm}$ and $B^{\rm I}_{\pm\pm}$ 
(having smallest masses of at least 0.034 and 0.053 eV, respectively). 
In addition, there is 
a different type of solution denoted $C^{\rm I}_{\pm}$, where the subscript 
denotes the sign of $\theta_{23} - \pi/4$. These two cases 
are special in the sense 
that they allow only a smallest mass between 0.004 and 0.013 eV. 
Example solutions are given in 
Table \ref{tab:Examples-of-numerical}. Note that some of the solutions with 
$\delta \to -\delta$ are connected by complex conjugation of the mass matrix. 

\begin{figure}[h]
\centering

\includegraphics[width=8cm]{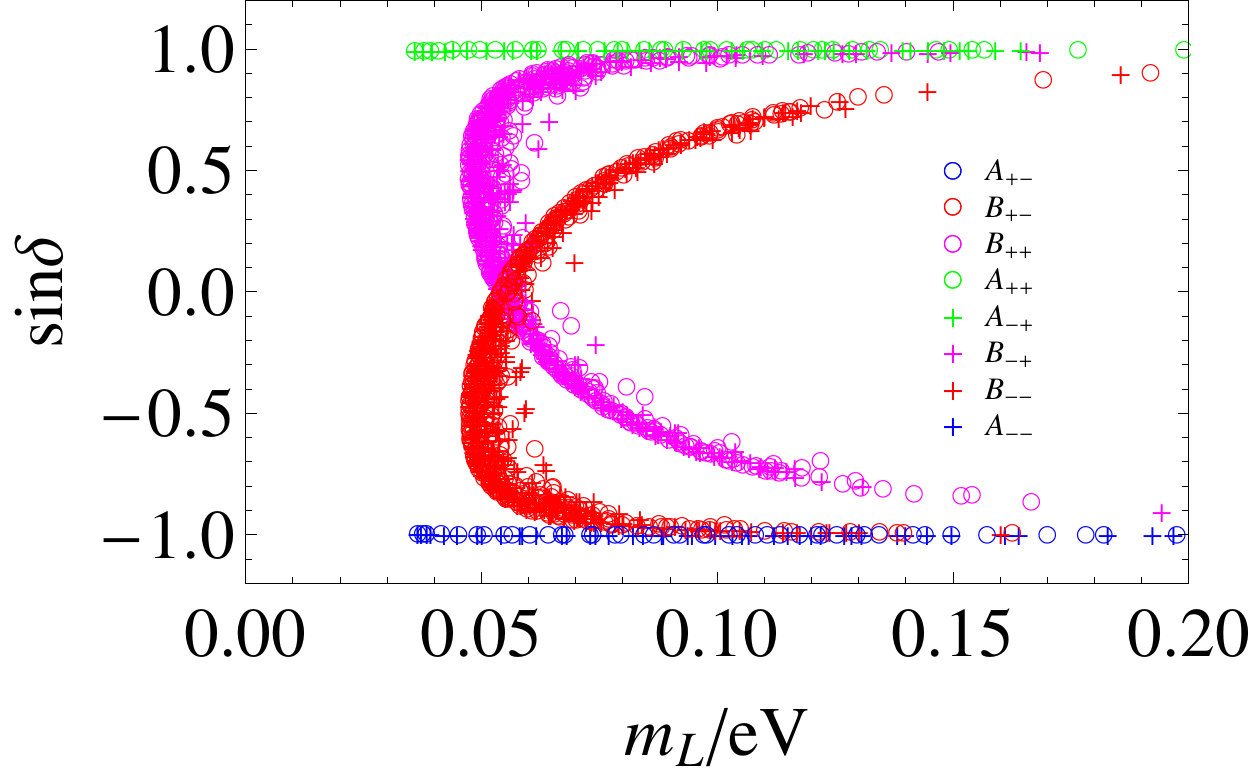}

\includegraphics[width=8cm]{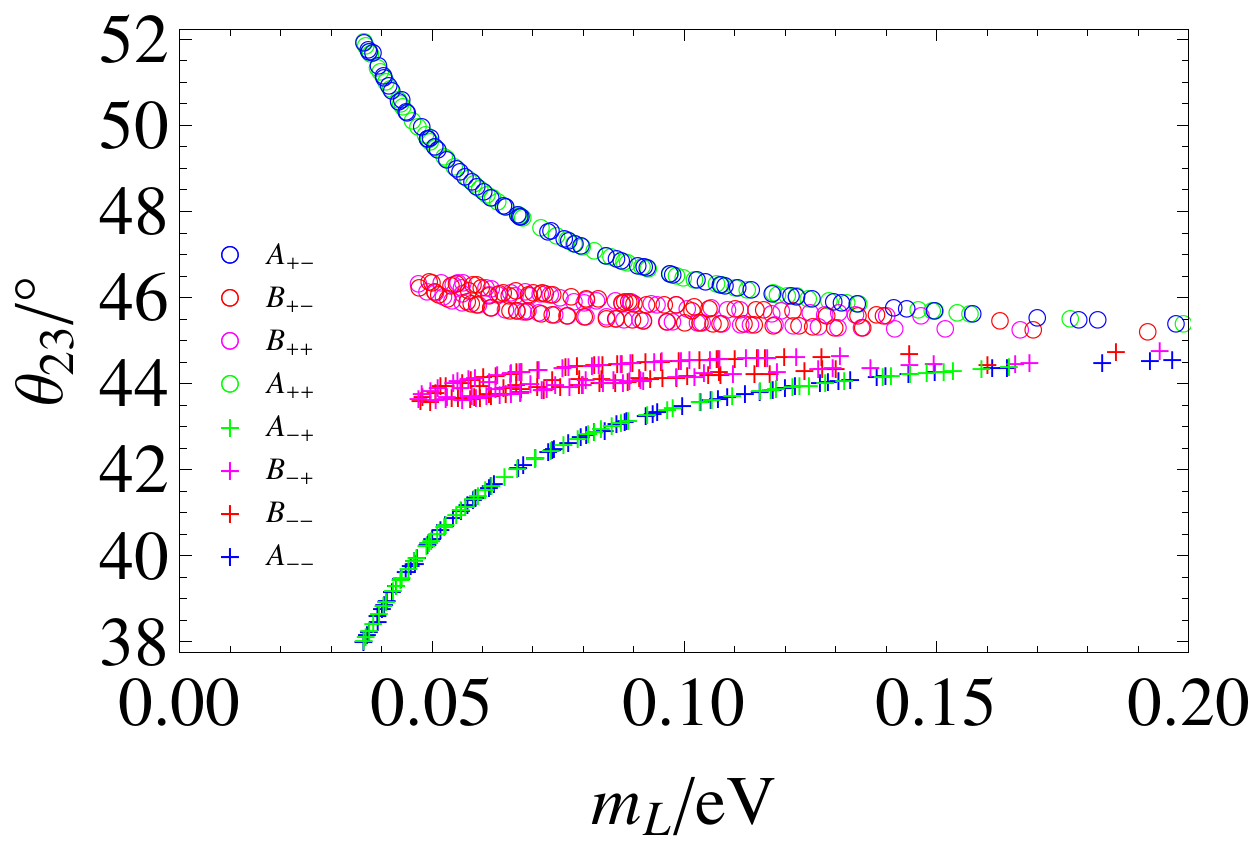}

\includegraphics[width=8cm]{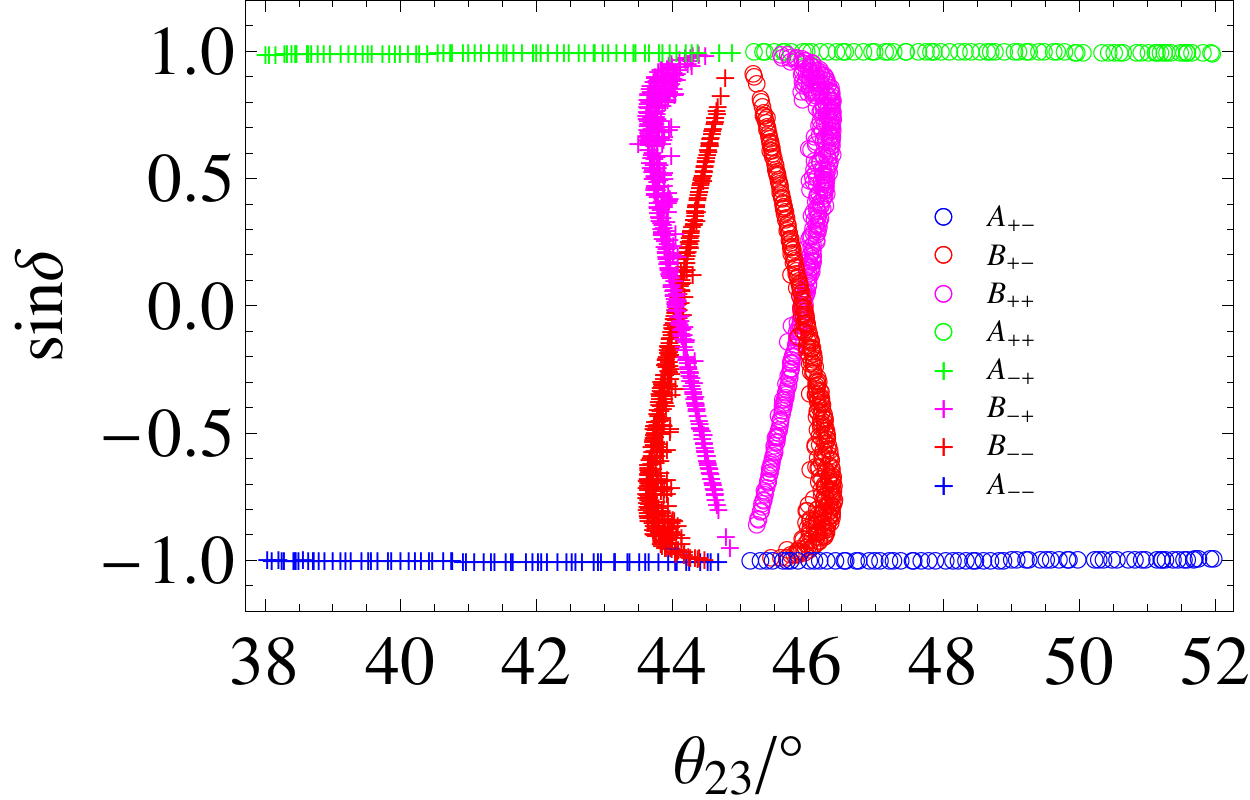}
\protect\caption{\label{fig:2n}Predicted relations from our model 
for the normal mass ordering. 
}
\end{figure}

\begin{figure}[h]
\centering

\includegraphics[width=8cm]{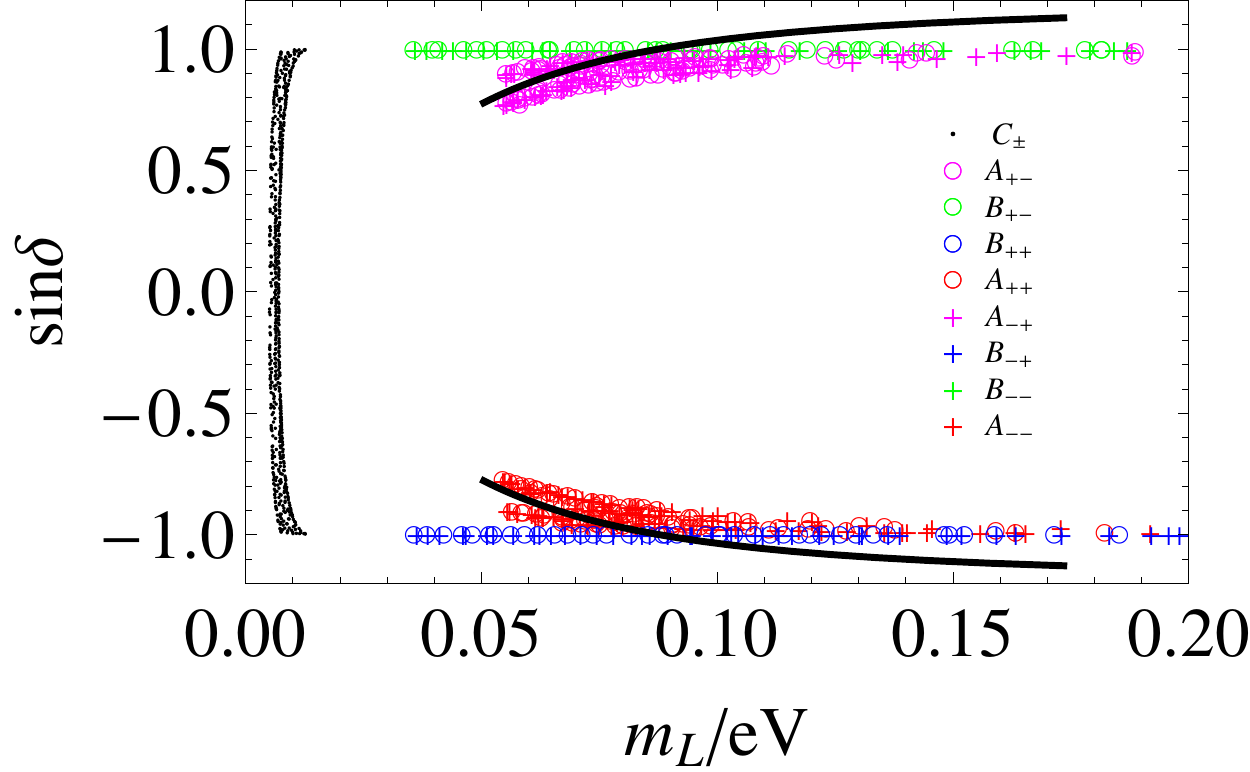}

\includegraphics[width=8cm]{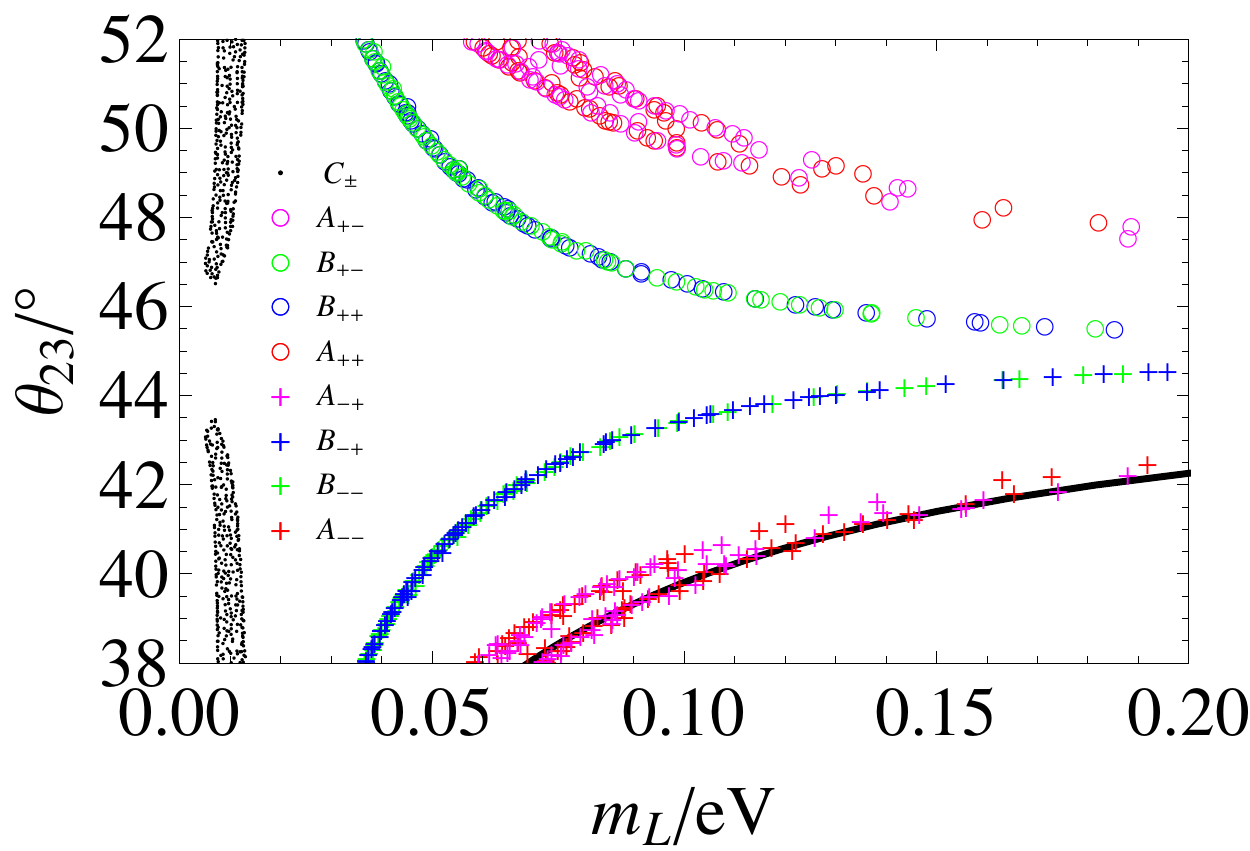}

\includegraphics[width=8cm]{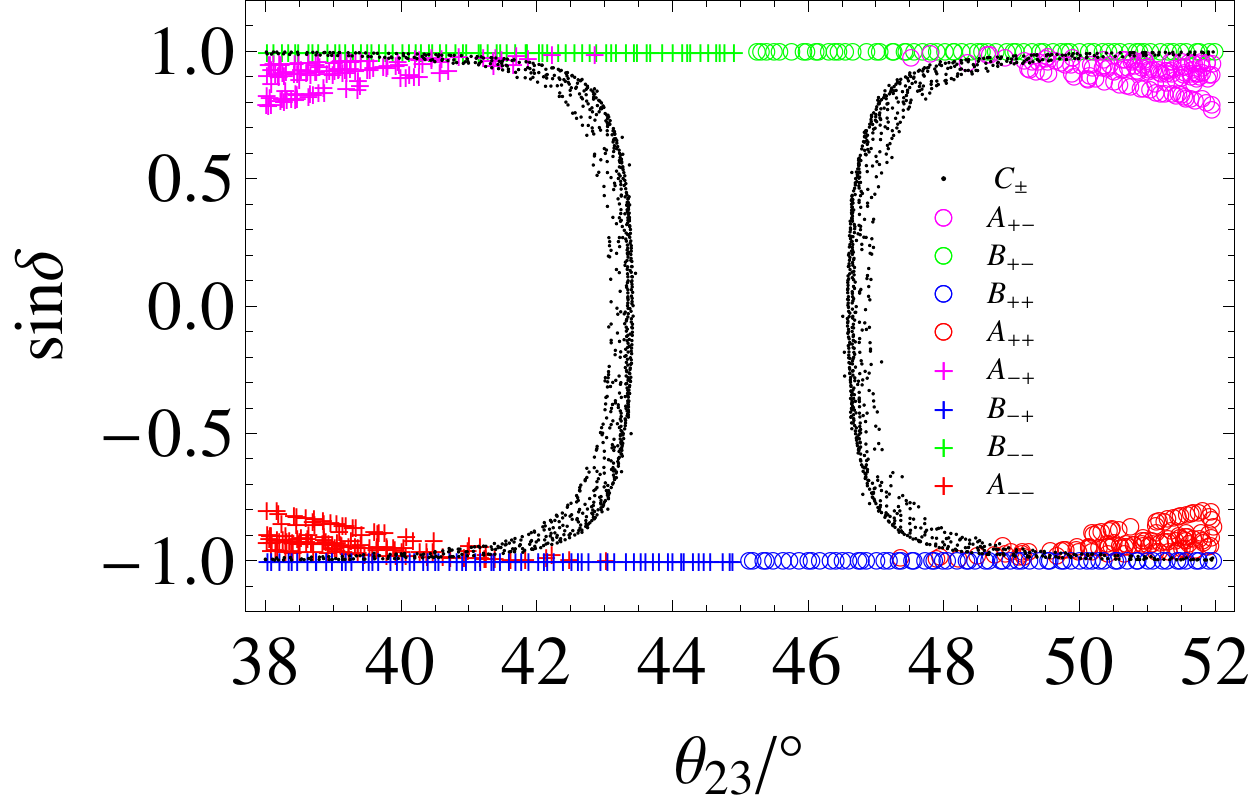}
\protect\caption{\label{fig:2i}
Predicted relations from our model 
for the inverted mass ordering. 
The black curves represent analytical results (see Eqs.\ 
(\ref{eq:0528-3}, \ref{eq:0528-6}) 
for explicit expressions) obtained from an approximate calculation, being 
well compatible with the numerical result.}
\end{figure}

\begin{figure}[h]
\centering

\includegraphics[width=8.1cm]{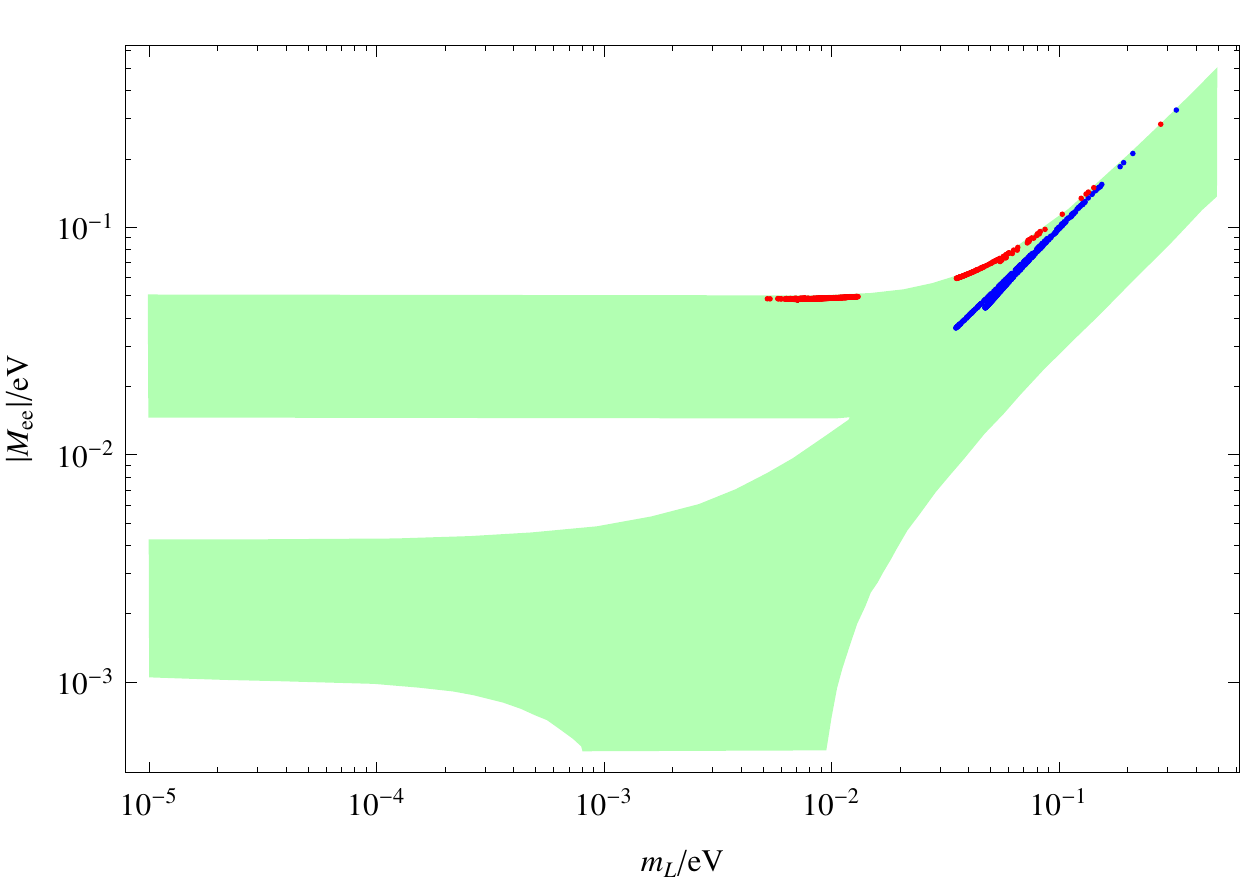}

\protect\caption{\label{fig:mass} The effective Majorana mass $M_{ee}$ in
neutrinoless double beta decay 
for both normal (blue) and inverted mass ordering (red, note the 
isolated red points corresponding to the $C^{\rm I}_{\pm}$ solutions). 
The green shaded area represents the currently allowed parameter space.  }
\end{figure}

The correlation between the interesting parameters 
$\theta_{23}$, $\delta$ and $m_L$ are given in Figs.\ \ref{fig:2n} 
and \ref{fig:2i}, respectively. 
Finally, Fig.\ \ref{fig:mass} summarizes the prediction 
of the model for neutrinoless double beta decay \cite{Rodejohann:2011mu}. 
We see in particular that for the inverted ordering 
it always holds that 
the effective mass takes essentially its largest possible values 
and that for the normal mass 
ordering the effective mass is non-zero. \\

\subsection{Analytical calculation}

Now we try to analytically find approximate expressions for one of the 
many possible solutions. From Table \ref{tab:Examples-of-numerical} we see 
that there are solutions with $r_2$ and $r_3$ close to one. 
Focusing on this case, we introduce the small parameters 
\begin{equation}
\delta_{2}\equiv r_{2}-1\,,\thinspace\delta_{3}\equiv r_{3}-1\label{eq:0527}
\end{equation}
in Eq.\ (\ref{eq:0512}). In this case the neutrino
mixing should be close to tri-bimaximal mixing (TBM) 
because if $\delta_{2},\delta_{3}=0$ the neutrino mixing is TBM. 
We further assume for simplicity that the neutrino
mass sum-rule as discussed at the end of Sec.\ \ref{sec:model} holds,
which is approximately true in this case as well. 

The deviation from TBM can be computed perturbatively under the assumption
$\delta_{2},\thinspace\delta_{3}\ll1$ and some details are found 
in the Appendix. The result turns out to be{\scriptsize{}
\begin{equation}
U\approx U_{{\rm TBM}}+\left[\begin{array}{ccc}
(\delta_{2}+\delta_{3})f_{11}(z) & (\delta_{2}+\delta_{3})f_{12}(z) & f_{13}(z)(\delta_{2}-\delta_{3})\\
. & . & f_{23}(z)(\delta_{2}-\delta_{3})\\
. & . & f_{33}(z)(\delta_{2}-\delta_{3})
\end{array}\right],\label{eq:0527-1}
\end{equation}
}where $z$ first appears in Eq.\ (\ref{eq:0512-15}) and the $f$-functions
are given in the appendix. The elements not given can be found there, 
but are not important here.  

The important point is that the deviations of $U_{e1}$ and $U_{e2}$
are proportional to $(\delta_{2}+\delta_{3})$ while the deviations
of $U_{e3}$, $U_{\mu 3}$ and $U_{\tau3}$ are proportional to 
$(\delta_{2}-\delta_{3})$. 
Note that $U_{e1}$ and $U_{e2}$ determine the value of $\theta_{12}$, 
which should not be too far away from the TBM value 
$\sin\theta_{12}=1/\sqrt{3}$. 
At the same time, $U_{e3}\propto\delta_{2}-\delta_{3}$ should 
be relatively large compared to the deviation of $\theta_{12}$. 
Thus, we simplify the analysis further by taking 
$\delta_{3}=-\delta_{2}$. Another assumption to make 
our life simpler is that $|1+z|\approx1$, which implies 
\[
z\approx e^{2i\alpha}-1\,.\label{eq:0527-5}
\]
The reason for this assumption is as follows: as mentioned above, 
we use that the actual mass spectrum $(m_{1},m_{2},m_{3})$ is still very 
close to the leading order one which is proportional to 
$(\frac{1}{1+z},1,\frac{-1}{1-z})$. With 
$\delta m^{2}/\Delta m^{2}\ll1$ it follows 
$(1-|\frac{1}{1+z}|^{2})/(1-|\frac{1}{1-z}|^{2})\ll1$
which implies $|1+z|$ should be very close to $1$. 
Note that if we assume $z\approx e^{2i\alpha}-1$, we are limited
to the inverted ordering because $|\frac{1}{1-z}|^{2}$ is always less than
$1$.

With the above assumptions (first taking $\delta_{3}=-\delta_{2}$
and then $z\approx e^{2i\alpha}-1$), Eq.\ (\ref{eq:0527-1}) can be
simplified to 
\begin{equation}
U\approx\left(\begin{array}{ccc}
\sqrt{\frac{2}{3}} & \frac{1}{\sqrt{3}} & g_{13}(\alpha)\delta_{2}\\
. & . & g_{23}(\alpha)\delta_{2}+\frac{1}{\sqrt{2}}\\
. & . & g_{33}(\alpha)\delta_{2}+\frac{1}{\sqrt{2}}
\end{array}\right),\label{eq:0528}
\end{equation}
where
\begin{equation}
g_{13,23,33}(\alpha)\equiv\frac{3-i\cot\alpha}{3\sqrt{2}},\frac{-3+2i\cot\alpha}{3\sqrt{2}},\frac{3-2i\cot\alpha}{3\sqrt{2}}.\label{eq:0528-1}
\end{equation}
Note that with $|g_{13}(\alpha)|\delta_{2}=\sin\theta_{13}$ we can replace
$\delta_{2}$ with $s_{13}/|g_{13}(\alpha)|$ and then extract 
$\tan\theta_{23}$ and $\sin\delta$ from Eq.\ (\ref{eq:0528}). They
can be expressed in terms of $\theta_{13}$ and $\alpha$, 
\begin{equation}
\begin{array}{c}
\tan\theta_{23}\approx\left|\frac{-2s_{13}z_{2}+i\sqrt{2}|z_{1}|}{2s_{13}z_{2}+i\sqrt{2}|z_{1}|}\right|,
\\
\sin\delta\approx\textrm{Im}\left[\frac{z_{1}\left(\sqrt{2}iz_{2}s_{13}+|z_{1}|\right)\left(\sqrt{2}z_{3}s_{13}+2i|z_{1}|\right)}{2|z_{1}|^{3}}\right],\label{eq:0528-3}
\end{array}
\end{equation}
where
\begin{eqnarray*}
z_{1} & \equiv & \cot\alpha-3i\,,\\
z_{2} & \equiv & 2\cot\alpha+3i\,,\\
z_{3} & \equiv & 3\cot\alpha+3i\,.
\end{eqnarray*}
Now we have derived approximate expressions for $\theta_{23}$ and
$\delta$ in terms of $\alpha$. The value of $\alpha$ can be related
to the lightest neutrino mass $m_{L}$ via 
\begin{equation}
m_{L}(\alpha)\approx\sqrt{\frac{1}{8}\Delta m^{2}\csc^{2}\alpha}\label{eq:0528-6}
\end{equation}
because the mass spectrum is $(m_{1}^{2},m_{2}^{2},m_{3}^{2})\approx m^{2}(1,1,|1+z|^{-2})$
in our approximation. From Eq.\ (\ref{eq:0528-3}) we can extract the 
following limit
\begin{equation}
\lim_{\alpha\rightarrow0}\tan\theta_{23}=\lim_{\alpha\rightarrow0}1-\frac{6\sqrt{2}s_{13}}{8s_{13}^{2}+1}|\tan\alpha|=1\,,\label{eq:0528-4}
\end{equation}
which implies $\alpha$ should be a small angle to make $\theta_{23}$
close to $45^{\circ}$. The limits of Eqs.\ (\ref{eq:0528-3}, \ref{eq:0528-6}) 
can also be computed, resulting in 
\begin{equation}
\lim_{\alpha\rightarrow0^{\pm}}\sin\delta=\pm(3\cos2\theta_{13}-4)\approx\pm1\label{eq:0528-5}
\end{equation}
and 
\begin{equation}
\lim_{\alpha\rightarrow0}m_{L}(\alpha)\propto\frac{\sqrt{\Delta m^{2}}}{|\alpha|}\,.
\label{eq:0528-5-1}
\end{equation}
Note the limit of $|\sin\delta|$ in Eq.\ (\ref{eq:0528-5}) is larger
than $1$ since 
$4-3\cos2\theta_{13}=1+6\theta_{13}^{2}+\mathcal{O}(\theta_{13}^{3})$.
This is due to the inaccuracy of our approximate calculation
where we omit all second-order corrections of $\delta_{2}$. The limit
(\ref{eq:0528-5-1}) implies that  small $\alpha$ results in  large
$m_{L}$, i.e.\ $m_{L}\rightarrow\infty\Longleftrightarrow\alpha\rightarrow0$. 
Therefore, for small $|\theta_{23}-45^{\circ}|$ the smallest mass 
$m_{L}$ is large. Furthermore, the larger $\alpha$ the larger is 
the deviation of $\theta_{23}$ from $\pi/4$, which means that 
there should be a lower bound on $m_{L}$. 
The above expressions also imply that 
$\lim_{m_{L}\rightarrow\infty}|\sin\delta|=1$, i.e.\ as neutrino mass increases 
the CP phase approaches one of its maximal values. 

Those features can be identified from the plots in Fig.\ 
\ref{fig:2i}, showing the accurateness of the analytical study.

\subsection{\label{sec:pheno}Phenomenological summary}
Let us summarize the phenomenological consequences of the model. 

First of all, the lightest neutrino mass cannot be zero or too small, 
quantitatively summarized from the previous results as follows: 
\begin{eqnarray}\nonumber
{\rm normal:}\thinspace & m_{L}\gs & 0.034 ~\textrm{eV,}\label{eq:0611}\\\nonumber
{\rm inverted:}\thinspace & m_{L}\in & (0.004,0.013) \,{\rm or}\, m_{L}\gs 
0.034~\textrm{eV.}\label{eq:0611-1}
\end{eqnarray}
The lower bound of $m_{L}$ for normal ordering has an important 
implication as the effective mass $M_{ee}$ is always non-zero. 
One finds 
\begin{eqnarray}\nonumber
{\rm normal:}\thinspace & M_{ee}\gs & 0.036~\textrm{eV,}\label{eq:0613}\\ \nonumber
{\rm inverted:}\thinspace & M_{ee}\in & (0.0482,0.0493) \,{\rm or}\, M_{ee}\gs 0.059~\textrm{eV.}\label{eq:0613-1}
\end{eqnarray}
We also note that for large $m_{L}$ ($\gtrsim0.1\,\,{\rm eV}$), 
\begin{equation}
m_{L}\approx M_{ee}\label{eq:0611-2}\,.
\end{equation}
This is due to the (approximately valid) sum-rule $2m_{2}^{-1}+m_{3}^{-1}=m_{1}^{-1}$ which 
will give the above relation for a quasi-degenerate 
spectrum \cite{Barry:2010yk}.

Another important feature of our model is the maximal CP violation.
As we can see from the top plots in Figs.\ \ref{fig:2n} 
and \ref{fig:2i}, both the $A^{\rm N}$ and $A^{\rm I}$ types of solution 
(green and blue points) always have maximal $|\sin\delta|$ with 
very little uncertainties. For the 
$B^{\rm N}$ and $B^{\rm I}$ types of solution, 
if $m_{L}$ is large enough, $|\sin\delta|$ also approaches its  
maximal value. This can be understood e.g.\ from our previous analytic 
computation which gives $\lim_{m_{L}\rightarrow\infty}|\sin\delta|=1$. 

The $C^{\rm I}$ solution in general do not have maximal CP 
violation. However, from the lower plot in Fig.\ \ref{fig:2i} we see that 
$\delta$ and $\theta_{23}$ are strongly correlated (black dots). 
If $\theta_{23}$ turns out to deviate significantly from $45^{\circ}$ 
such as $\theta_{23}<42^{\circ}$ or $\theta_{23}>48^{\circ}$, then 
the $C^{\rm I}$ solutions also predict maximal $|\sin\delta|$. 

The two bottom plots in Figs.\ \ref{fig:2n} and \ref{fig:2i} show that
if large $|\theta_{23}-45^{\circ}|$ is observed in the future,
then $|\sin\delta|$ must be close to its maximal value. For the inverted 
ordering this requires $|\theta_{23}-45^{\circ}|\gtrsim3^{\circ}$, as 
just discussed, while for the normal ordering 
it requires $|\theta_{23}-45^{\circ}|\gtrsim1.5^{\circ}$. It 
is interesting to note that such a deviation of $\theta_{23}$ and
a maximal $|\sin\delta|$ are simultaneously (still rather mildly) 
preferred by current
global fit as the best-fit of $(\theta_{23},\sin\delta)$ is $(41.4^{\circ},-0.94)$
for normal ordering 
and $(42.4^{\circ},-0.83)$ for inverted ordering \cite{NeuFitLisi}. 

Finally, since in the large $m_{L}$ limit $\theta_{23}$ goes to $45^{\circ}$,
a significant deviation of $\theta_{23}$ from $45^{\circ}$ 
implies an upper bound on $m_{L}$. For example 
if $|\theta_{23}-45^{\circ}|\gtrsim3^{\circ}$ in 
the normal ordering 
then from Fig.\ \ref{fig:2n} we get $m_{L}\lesssim 0.06$ eV, which
constrains $m_{L}$ to a very narrow region $(0.034,0.06)$ eV. 

It is also possible to rule out a mass ordering in this model due
to the different structures of solutions. For example,
if the future bound 
on $m_{L}$ is pushed below $0.034$ eV, then only the $C^{\rm I}$ solutions 
survive. Also, since $(\theta_{23},\delta)$ shown in the bottom 
plots in Figs.\ \ref{fig:2n}, \ref{fig:2i} have very different 
distributions for both possible mass orderings, 
it is also possible to distinguish them with precise measurements  
on $\theta_{23}$ and $\delta$. 

In summary, if $m_{L}$ is large, we have clear predictions on $\delta$,
$\theta_{23}$ and $M_{ee}$, which should be close to their large $m_{L}$
limit
\begin{equation}
\lim_{m_{L}\rightarrow\infty}(|\sin\delta|,\theta_{23},M_{ee})=(1,45^{\circ},m_{L})\,.\label{eq:0612}
\end{equation}
If $m_{L}$ is small, then we have some more interesting predictions
among these parameters, such as large deviations from $\theta_{23}=45^{\circ}$, 
correlations between $\theta_{23}$ and $\delta$ as well as 
with the mass ordering.

\fi

\section{\label{sec:Conclusion}Conclusion}

We presented in this model a flavor symmetry model based on $A_4$ within a left-right 
symmetric framework. Various aspect exist that make this environment different from the 
usual model building. This includes the necessity to treat the particles in left- 
and right-handed doublets, but more crucially the fact that residual symmetries 
from breaking the full flavor group do not make it in the mass matrices and hence 
do not determine the mixing. Furthermore, the discrete left-right symmetry should be 
parity rather than charge conjugation, in order to avoid inconsistencies between the 
flavor and charge conjugation symmetries. 

Taking all this into account, we were discussing a left-right symmetric model with 
$A_4$ flavor symmetry and analyzed its predictions. No flavor changing neutral currents from the Higgs bi-doublet are present.    
Several distinct solutions for the neutrino sector 
were possible, many of which prefering maximal CP violation as currently 
prefered by data. Various other predictions and correlations exist which would 
allow for tests of the model. 

The various constraints that left-right symmetric theories impose on 
flavor symmetry models will allow for further analyses, both conceptual 
as well as phenomenological. The possibility to use left-right symmetry as a first bottom-up 
step to approach GUT flavor symmetries is another attractive option to study. Such endevours 
will be left for future studies.

\begin{acknowledgments}
We thank Sudhanwa Patra for useful discussion on left-right symmetry.
WR is supported by the Max Planck Society in the project MANITOP, and
by the DFG in the Heisenberg programme with grant RO 2516/6-1. 
XJX by the China Scholarship Council (CSC).
\end{acknowledgments}

\bibliographystyle{apsrev4-1}
\bibliography{ref}

\appendix

\section{Approximate diagonalization of Majorana mass matrices}

\subsection{General formulae}

If a Majorana mass matrix can be written as 
\begin{equation}
M=M_{0}+\delta M\,,\label{eq:0529-1}
\end{equation}
where $\delta M\ll M_{0}$ and $M_{0}$ can be diagonalized by $U_{0}$
\begin{equation}
U_{0}^{T} M_{0} U_{0}={\rm diag}(m_{1},m_{2},m_{3})\,,\label{eq:0529-2}
\end{equation}
then $M$ can be approximately diagonalized by $U$
\begin{equation}
U^{T} M U\approx{\rm diag}(m_{1}+\delta m_{1},m_{2}+\delta m_{2},m_{3}+\delta m_{3})\,,\label{eq:0529-3}
\end{equation}
where $U$ can be computed as 
\begin{equation}
\begin{array}{c}
U\approx U_{0}(1+iT) \,,
\\
T_{ij}=\frac{i(A_{ij}m_{i}^{*}+A_{ij}^{*}m_{j})}{|m_{i}|^{2}-|m_{j}|^{2}}\thinspace(i\neq j),\thinspace T_{ii}=0\,,\\
A\equiv U_{0}^{T} \delta M U_{0} \,,
\,\,\,\delta m_{i}=A_{ii}\,.\label{eq:0529-6}
\end{array}
\end{equation}
The above formulae only hold for $T_{ij}\ll1$, which requires not
only $\delta M\ll M_{0}$ but also $|m_{i}|^{2}-|m_{j}|^{2}$ are not
very small, i.e.\ degeneracies in the zeroth-order mass spectrum would
invalidate the approximate diagonalization. 

These formulae can be derived as follow. 
Note that 
\begin{equation}
U_{0}^{T} M U_{0}=\left(\begin{array}{ccc}
m_{1}\\
 & m_{2}\\
 &  & m_{3}
\end{array}\right)+A\label{eq:0529-7}
\end{equation}
is quasi-diagonal and we need to perform a small rotation $1+iT$
where $T=T^{\dagger}$ and $T\ll1$ to diagonalize it: 
\begin{equation}
\left(\begin{array}{ccc}
m_{1}\\
 & m_{2}\\
 &  & m_{3}
\end{array}\right)+A=(1-iT)^{T}\left(\begin{array}{ccc}
\tilde{m}_{1}\\
 & \tilde{m}_{2}\\
 &  & \tilde{m}_{3}
\end{array}\right)(1-iT).\label{eq:0529-8}
\end{equation}
We can assume $T_{ii}=0$ because if we rephase each column of $U$,
i.e.\ $U\rightarrow U{\rm diag}(e^{i\alpha_{1}},e^{i\alpha_{2}},e^{i\alpha_{3}})$,
$U$ still can diagonalize $M$ and any non-zero $T_{ii}$ can be
absorbed into such rephasing.

The next-to-leading order in Eq.\ (\ref{eq:0529-8}) gives 
\[
A\approx-i\left(\begin{array}{ccc}
0 & t_{12} & t_{13}\\
. & 0 & t_{23}\\
. & . & 0
\end{array}\right)+\left(\begin{array}{ccc}
\delta m_{1}\\
 & \delta m_{2}\\
 &  & \delta m_{3}
\end{array}\right),
\]
where 
\[
t_{ij}\equiv T_{ji}m_{j}+T_{ij}m_{i}\,~,\delta m_{i}=\tilde{m_{i}}-m_{i}\,.
\]
So we have 
\[
A_{ij}=-i(T_{ji}m_{j}+T_{ij}m_{i})
\]
and its conjugate
\[
A_{ij}^{*}=i(T_{ij}m_{j}^{*}+T_{ji}m_{i}^{*})\,.
\]
Now we can solve the above two equations with respect to $T_{ij}$
and $T_{ji}$ to get Eq.\ (\ref{eq:0529-6}).

\subsection{Diagonalization of the neutrino mass matrix in the model}

We use the general formulae to diagonalize Eq.\ (\ref{eq:0512}). First we compute $A$ defined
in Eq.\ (\ref{eq:0529-6}) 
\[
A=m\left(\begin{array}{ccc}
\frac{\delta_{2}+\delta_{3}}{3(z+1)} & -\frac{(z+2)\left(\delta_{2}+\delta_{3}\right)}{3\sqrt{2}(z+1)} & \frac{z\left(\delta_{3}-\delta_{2}\right)}{\sqrt{3}\left(z^{2}-1\right)}\\
-\frac{(z+2)\left(\delta_{2}+\delta_{3}\right)}{3\sqrt{2}(z+1)} & \frac{2}{3}\left(\delta_{2}+\delta_{3}\right) & \frac{z\left(\delta_{2}-\delta_{3}\right)}{\sqrt{6}(z-1)}\\
\frac{z\left(\delta_{3}-\delta_{2}\right)}{\sqrt{3}\left(z^{2}-1\right)} & \frac{z\left(\delta_{2}-\delta_{3}\right)}{\sqrt{6}(z-1)} & \frac{\delta_{2}+\delta_{3}}{z-1}
\end{array}\right),
\]
and then $T$ from Eq.\ (\ref{eq:0529-6}) 
\[
T=\left(\begin{array}{ccc}
0 & \frac{i(3z+(z+1)z^{*}+4)\left(\delta_{2}+\delta_{3}\right)}{3\sqrt{2}\left(|z|^{2}+z+z^{*}\right)} & \frac{i(z(z^{*}-1)+\Re(z))\left(\delta_{2}-\delta_{3}\right)}{2\sqrt{3}\Re(z)}\\
. & 0 & \frac{(izz^{*}+2\Im(z))\left(\delta_{2}-\delta_{3}\right)}{\sqrt{6}\left(|z|^{2}-2\Re(z)\right)}\\
. & . & 0
\end{array}\right).
\]
So from $U$ in Eq.\ (\ref{eq:0529-6}) we get 
\[
U_{e1}=\frac{\frac{(z+(z+3)z^{*}+4)\left(\delta_{2}+\delta_{3}\right)}{|z|^{2}+z+z^{*}}+6}{3\sqrt{6}}
\]
\[
U_{e2}=\frac{3-\frac{(3z+(z+1)z^{*}+4)\left(\delta_{2}+\delta_{3}\right)}{|z|^{2}+z+z^{*}}}{3\sqrt{3}}
\]
\[
U_{e3}=-\frac{z|z|^{2}(z^{*}-1)\left(\delta_{2}-\delta_{3}\right)}{3\sqrt{2}\left(|z|^{2}-2\Re(z)\right)\Re(z)}
\]
\[
U_{\mu 1}=\frac{\frac{3\left(|z|^{2}+z-\Re(z)\right)\left(\delta_{2}-\delta_{3}\right)}{\Re(z)}+\frac{2(z+(z+3)z^{*}+4)\left(\delta_{2}+\delta_{3}\right)}{|z|^{2}+z+z^{*}}-6}{6\sqrt{6}}
\]
\[
U_{\mu 2}=\frac{\frac{3(z+(z-1)z^{*})\left(\delta_{2}-\delta_{3}\right)}{|z|^{2}-2\Re(z)}+\frac{(3z+(z+1)z^{*}+4)\left(\delta_{2}+\delta_{3}\right)}{|z|^{2}+z+z^{*}}+6}{6\sqrt{3}}
\]
\[
U_{\mu 3}=\frac{\frac{(z(z^{*}-1)+\Re(z))\left(\delta_{2}-\delta_{3}\right)}{\Re(z)}-\frac{2\left(|z|^{2}-z+z^{*}\right)\left(\delta_{2}-\delta_{3}\right)}{|z|^{2}-2\Re(z)}+6}{6\sqrt{2}}
\]
\[
U_{\tau 1}=\frac{\frac{3\left(|z|^{2}+z-\Re(z)\right)\left(\delta_{2}-\delta_{3}\right)}{\Re(z)}-\frac{2(z+(z+3)z^{*}+4)\left(\delta_{2}+\delta_{3}\right)}{|z|^{2}+z+z^{*}}+6}{6\sqrt{6}}
\]
\[
U_{\tau 2}=\frac{\frac{3(z+(z-1)z^{*})\left(\delta_{2}-\delta_{3}\right)}{|z|^{2}-2\Re(z)}-\frac{(3z+(z+1)z^{*}+4)\left(\delta_{2}+\delta_{3}\right)}{|z|^{2}+z+z^{*}}-6}{6\sqrt{3}}
\]
\[
U_{\tau 3}=\frac{-\frac{(z(z^{*}-1)+\Re(z))\left(\delta_{2}-\delta_{3}\right)}{\Re(z)}+\frac{2\left(|z|^{2}-z+z^{*}\right)\left(\delta_{2}-\delta_{3}\right)}{|z|^{2}-2\Re(z)}+6}{6\sqrt{2}}
\]
The $f$-functions used in Eq.\ (\ref{eq:0527-1}) can be extract
from above results, for example,
\begin{equation}
f_{11}(z)\equiv\frac{(z+3)z^{*}+z+4}{3\sqrt{6}\left(|z|^{2}+z^{*}+z\right)},\label{eq:0527-2}
\end{equation}
\begin{equation}
f_{12}(z)\equiv-\frac{(z+1)z^{*}+3z+4}{3\sqrt{3}\left(|z|^{2}+z^{*}+z\right)},\label{eq:0527-2-1}
\end{equation}
\begin{equation}
f_{13}(z)\equiv-\frac{z|z|^{2}(z^{*}-1)}{3\sqrt{2}\Re(z)\left(|z|^{2}-2\Re(z)\right)}.\label{eq:0527-2-2}
\end{equation}

\end{document}